# Conformally Invariant Vector-Tensor Field Theories Consistent with Conservation of Charge in a Four-Dimensional Space

by

Gregory W. Horndeski
2814 Calle Dulcinea
Santa Fe, NM 87505-6425

e-mail:
horndeskimath@gmail.com

November 4, 2017

# ABSTRACT

In this paper I shall construct, in a four-dimensional space, all vector-tensor field theories that are conformally invariant, consistent with conservation of charge, and flat space compatible. By the last assumption I mean that the Lagrangian of the theory in question, is well-defined and differentiable, when evaluated for either a flat metric tensor (and) or vanishing vector field. Under these assumptions there exists only two classes of Lagrangians. One is represented by the Lagrangian which yields the Bach tensor density multiplied by a constant, while the other is represented by a constant multiple of the usual Lagrangian that yields Maxwell's equation. Thus, under the aforementioned assumptions, the field equation obtained by varying the vector field, must be Maxwell's.

# DEDICATION

This paper is dedicated to the memory of my wife, Sharon Winklhofer Horndeski, Ph.D., J.D., who died on August 8, 2017, while I was preparing this manuscript. I shall miss her greatly.

**Section 1: Introduction**

In Einstein's theory of gravity and electromagnetism, the field equations governing the Lorentzian metric tensor, $g_{ab}$, and the vector potential, $\psi_a$, in regions devoid of sources, are

$$G^{ab} - T_M{}^{ab} = 0 , \qquad \text{Eq.1.1}$$

and

$$F^{ab}{}_{|b} = 0 , \qquad \text{Eq.1.2}$$

where

$$T_M{}^{ab} := 2[F^{ac} F^b{}_c - \tfrac{1}{4} g^{ab} F^{cd} F_{cd} ] \qquad \text{Eq.1.3}$$

and

$$F_{ab} := \psi_{a,b} - \psi_{b,a} . \qquad \text{Eq.1.4}$$

The notation used in this paper is the same as that employed in [1], except where stipulated to the contrary. Eqs. 1.1 and 1.2 are referred to as the source-free Einstein-Maxwell field equations, and they are derivable from the Einstein-Maxwell Lagrangian

$$L_{EM} := g^{½}R - g^{½} F^{ab} F_{ab} . \qquad \text{Eq.1.5}$$

In the presence of sources for the gravitational and electromagnetic field, Eqs.1.1 and 1.2 are modified by the addition of $8\pi T^{ab}$ and $-4\pi J^a$ to the right-hand sides of these equations respectively. Here $T^{ab}$ and $J^a$ denote the energy-momentum tensor and charge-current vector for the source fields. In general, the requirement that charge be conserved translates to the demand that $J^a{}_{|a} = 0$. Since $F^{ab}{}_{|ba} \equiv 0$, we see that in the presences of sources, the Einstein-Maxwell field equations are compatible with charge conservation.

Another interesting aspect of these equations is that the Lagrangian which yields $g^{½}T_M^{ab}$, and the electromagnetic field equations; *viz.,*

$$L_M := -g^{½} F^{ab} F_{ab} , \qquad \text{Eq.1.6}$$

is invariant under a conformal transformation. This in turn implies that $E^a{}_b(L_M)$ and $E^a(L_M)$ are conformally invariant. So $L_M$ generates field equations which are conformally invariant and compatible with charge conservation. The natural question to ask is whether $L_M$ is unique in that regard? To resolve that problem I first have to introduce some terminology.

We shall say that a physical field theory is a vector-tensor field theory, if the field variables are the components of a covariant vector field, $\psi_a$, and a metric tensor field, $g_{ab}$, with the field equations being derivable from a Lagrangian of the form

$$L = L(g_{ab}; g_{ab,c} ; \ldots ; \psi_a ; \psi_{a,b} ; \ldots) , \qquad \text{Eq.1.7}$$

which is of finite differential order in the field variables, and assumed to be a scalar density. We define the Euler-Lagrange tensor densities by

$$E^{ab}(L) := -\frac{\partial L}{\partial g_{ab}} + \frac{d}{dx^c} \frac{\partial L}{\partial g_{ab,c}} + \ldots \qquad \text{Eq.1.8}$$

and

$$E^a(L) := -\frac{\partial L}{\partial \psi_a} + \frac{d}{dx^b} \frac{\partial L}{\partial \psi_{a,b}} + \ldots . \qquad \text{Eq.1.9}$$

The field theory will be said to be of $k^{th}$ order, if one of the set of field tensor densities has derivatives of $k^{th}$ order, in a least one of the field variables. Note that the Euler-Lagrange tensor densities defined in Eqs.1.8 and 1.9 are the negative of those used in [1].

Let L be the Lagrangian of a vector-tensor field theory. Under the conformal transformation $g_{ab} \rightarrow g'_{ab} := e^{2\sigma} g_{ab}$, where $\sigma$ is a differentiable real valued scalar field, L generates a Lagrangian L', defined by

$$L'(g'_{ab}; g'_{ab,c}; \ldots; \psi_a; \psi_{a,b}; \ldots) := L(g'_{ab}; g'_{ab,c}; \ldots; \psi_a; \psi_{a,b}; \ldots).$$

L is said to be conformally invariant if $L' = L$, when $g'_{ab}$ is replaced throughout L' by $e^{2\sigma} g_{ab}$.

A vector-tensor field theory will be said to be conformally invariant if $E^a{}_b(L)$, and $E^a(L)$, are conformally invariant. If L is conformally invariant, or conformally invariant up to a divergence, then it is easily demonstrated that its associated vector-tensor field theory will be conformally invariant.

We shall say that a vector-tensor field theory is consistent with charge conservation if the Euler-Lagrange tensor density $E^a(L)$, is identically divergence free. This guarantees that charge will be conserved, because in the presence of charge the vector field equation is $E^a(L) = -16\pi \, g^{\frac{1}{2}} J^a$, where $J^a$ is the charge-current vector. As pointed out above, the Einstein-Maxwell field equations are compatible with conservation of charge, although they are not conformally invariant.

Another thing we note about the Einstein-Maxwell field equations is that the Lagrangian of this theory is well defined, and differentiable (as a tensorial concomitant) when evaluated for a flat metric tensor, and vanishing vector potential. With that in mind I shall define a vector-tensor field theory to be flat space compatible, if the Lagrangian of that theory is well defined and differentiable when evaluated for either a flat metric tensor (and) or vanishing vector potential. When this is the case the field tensors of that theory will also be well defined and differentiable when evaluated for either a flat metric tensor (and ) or vanishing vector potential. It seems physically reasonable to demand that a vector-tensor field theory be flat space compatible, since that will guarantee that the field equations of that theory do not blow up when nothing is in the space.

There exists a trivial example of a vector-tensor field theory, which is flat space compatible, conformally invariant, and consistent with charge conservation. This theory is generated by the pure metric Lagrangian,

$$L_B := -½ g^{½} C^{hijk} C_{hijk} , \qquad \text{Eq.1.10}$$

where $C^{hijk}$ is the Weyl tensor, which in a 4-dimensional space, is defined by

$$C^{hijk} := R^{hijk} + ½(g^{hk}R^{ij} + g^{ij}R^{hk} - g^{hj}R^{ik} - g^{ik}R^{hj}) + \tfrac{1}{6}R(g^{hj}g^{ik} - g^{hk}g^{ij}). \qquad \text{Eq.1.11}$$

The Euler-Lagrange tensor densities associated with $L_B$ are

$$E^{ab}(L_B) := g^{½}B^{ab} = g^{½}(C^{aijb}{}_{|ij} + C^{ajib}{}_{|ij} - R_h{}^a{}_{jk}C^{hbjk} + ¼g^{ab}C^{hijk}C_{hijk}) \qquad \text{Eq.1.12}$$

and

$$E^a(L_B) = 0 . \qquad \text{Eq.1.13}$$

(A derivation of Eq.1.12 can be found on page 9 of [2].) The tensor $B^{ab}$ presented in

Eq.1.12 is the well-known Bach tensor [3], which in a four-dimensional space can be expressed as follows:

$$B^{ab} = -\Box R^{ab} + \tfrac{1}{3} R^{|ab} + \tfrac{1}{6} g^{ab} \Box R + 2 R^{habk} R_{hk} + \tfrac{1}{2} g^{ab} R^{hk} R_{hk} + \tfrac{2}{3} R R^{ab} - \tfrac{1}{6} g^{ab} R^2 \ .$$

So right now we know of two conformally invariant, flat space compatible vector-tensor field theories that are consistent with charge conservation. The question is: Are there any more such theories? The answer is provided by the following

**Theorem:** In a four-dimensional space, let L be a Lagrangian which generates a conformally invariant, flat space compatible, vector-tensor field theory which is consistent with charge conservation. Then the Euler-Lagrange tensor densities associated with L can also be obtained from the Lagrangian $bL_B + \beta L_M$ for a suitable choice of the constants b and β. $L_M$ and $L_B$ are defined by Eqs.1.6 and 1.10. ∎

The above Theorem tells us something amazing about Maxwell's equations of electromagnetism in a curved space. It tells us that if we are looking for a flat space compatible, vector-tensor field theory in a space of four-dimensions, which is conformally invariant, and consistent with conservation of charge, then the electromagnetic field equations must be Maxwell's. However, the energy-momentum tensor of that theory can differ from Maxwell's by a term involving the Bach tensor. I shall now address that issue.

Let L be the Lagrangian of a vector-tensor field theory. We can write L as

$$L = L_1 + \ldots + L_\lambda ,$$

where $L_1, \ldots, L_\lambda$ are scalar densities, which can not be further simplified as the sum of other scalar densities. We shall say that the vector-tensor theory generated by L is a true vector-tensor field theory if each of the Lagrangians $L_1, \ldots, L_\lambda$ that comprise L, are such that at least one of their Euler-Lagrange tensor densities actually involve the vector potential, or its derivatives, in some way. Thus the Lagrangian $bL_B + \beta L_M$ describes a true vector-tensor field theory only if $b = 0$, and $\beta \neq 0$. Consequently the above Theorem immediately gives rise to the following

**Corollary:** In a four-dimensional space let L be a Lagrangian which generates a conformally invariant, flat space compatible, true vector-tensor field theory, which is consistent with charge conservation. Then the Euler-Lagrange tensor densities associated with L can also be obtained from $\beta L_M$, where $\beta$ is a constant and $L_M$ is the Maxwell Lagrangian, defined by Eq.1.6.∎

In the next section I shall provide a detailed proof of the Theorem. This will be followed in the concluding section, by remarks upon the Theorem, gauge-tensor field theories, and conformally invariant vector-tensor field theories.

Before I close this introduction, let me quickly sketch how the Theorem will be proved. I begin by showing that if L satisfies the assumptions of the Theorem, then $A^{ij} := E^{ij}(L)$ and $C^i := E^i(L)$, must be devoid of explicit dependence on the vector field. I then go on to show that $A^{ij}$ and $C^i$ are at most of fourth-order in $g_{ab}$ and third-order in $\psi_a$. Next I construct the general form of $C^i$, and show that $C^i = E^i(\beta L_M)$, for a suitable choice of the constant β. The Lagrangian $L := L - \beta L_M$ will satisfy the assumptions of the Theorem and be such that $E^i(L) = 0$. The proof ends by demonstrating that $L$ is equivalent to the Lagrangian $bL_B$ for a suitable choice of the constant b. Now for the details, which should be familiar to those who have read [2].

**Section 2: Proof of the Theorem**

The proof will consist of a sequence of Lemmas similar to those used in [2]. Since the signature of the metric tensor will not be significant in what we are about to do, I shall assume that it is arbitrary, but fixed.

The first lemma will provide us with an easy way to recognized conformally invariant vector-tensor field theories.

**Lemma 1:** Let L be the Lagrangian of a vector-tensor field theory. That field theory will be conformally invariant if and only if $E^{ab}(L)$ is trace-free. If $E^{ab}(L)$ is trace-free, then L is conformally invariant up to a divergence.

**Proof:** ⇒ The Euler-Lagrange tensor densities of a vector-tensor field theory are related by the identity (*see,* [1], or page 49 of [4])

$$E_a{}^b(L)_{|b} = -\tfrac{1}{2} F_{ab} E^b(L) - \tfrac{1}{2} \psi_a E^b(L)_{|b}. \qquad \text{Eq.2.1}$$

If $g'_{ab} := e^{2\sigma} g_{ab}$, we let $E_a{}^b(L)'$ and $E^a(L)'$ denote $E_a{}^b(L)$ and $E^a(L)$ built from $g'_{ab}$ and $\psi_a$. Since Eq.2.1 is an identity it is valid for every metric tensor and vector field. Thus

$$E_a{}^b(L)'_{|'b} = -\tfrac{1}{2}F_{ab}E^b(L)' - \tfrac{1}{2}\psi_a E^b(L)'_{|'b},  \qquad \text{Eq.2.2}$$

where "$_|$" denotes covariant differentiation with respect to the Levi-Civita connection of $g'_{ab}$. Due to our assumption of conformal invariance $E^a(L) = E^a(L)'$. Thus the right-hand sides of Eqs.2.1 and 2.2 are equal. Consequently

$$E_a{}^b(L)'_{|'b} = E_a{}^b(L)_{|b}. \qquad \text{Eq.2.3}$$

Since, by assumption, $E_a{}^b(L) = E_a{}^b(L)'$, we can use the fact that

$$\Gamma'^r{}_{st} = \Gamma^r{}_{st} + \sigma_{,s}\delta^r{}_t + \sigma_{,t}\delta^r{}_s - g_{st}g^{rm}\sigma_{,m}$$

in Eq.2.3, to deduce that $E_a{}^a(L) = 0$.

$\Leftarrow$ Let $g(t)_{ab} := (1-t)g_{ab} + t\, g'_{ab}$, $0 \le t \le 1$, denote the convex combination of $g_{ab}$ and $g'_{ab}$. So $g(t)_{ab} = (1-t+te^{2\sigma})g_{ab}$, is a pseudo-Riemannian metric with the same signature as $g_{ab}$. We now define a one parameter family of Lagrangians, $L(t)$ by

$$L(t) := L(g(t)_{ab}; g(t)_{ab,c}; \ldots; \psi_a; \psi_{a,b}; \ldots) .$$

If we let $E^{ab}(L(t))$ denote $E^{ab}(L)$ evaluated for $g(t)_{ab}$ and $\psi_a$, then it is a straight forward matter to demonstrate that

$$\frac{dL(t)}{dt} = E^{ab}(L(t))\frac{dg(t)_{ab}}{dt} + \frac{d}{dx^i}V(t)^i \qquad \text{Eq.2.4}$$

where $V(t)^i$ is a contravariant vector density built from $t$, $\sigma$, $g_{ab}$ and $\psi_a$. Since $E^{ab}(L)$ is trace free, we know that

$$0 = E^{ab}(L(t))\, g(t)_{ab} = E^{ab}(L(t))\, g_{ab}\,(1-t+te^{2\sigma}) ,$$

and thus

$$E^{ab}(L(t)) \, g_{ab} = 0 \, .$$

Consequently Eq.2.4 implies that

$$\frac{dL(t)}{dt} = \frac{d}{dx^i} V(t)^i \, . \qquad \text{Eq.2.5}$$

If we integrate Eq.2.5 with respect to t, from 0 to 1, we find that

$$L(1) - L(0) = \text{a divergence.}$$

But $L(1) = L'$ and $L(0) = L$, and so if $E^{ab}(L)$ is trace free, L is conformally invariant up to a divergence. This in turn implies that $E^a{}_b(L)$ and $E^a(L)$ are conformally invariant. ∎

Now that we have found an easy way to spot conformally invariant vector-tensor field theories, what we need is an equally simple way to tell if a vector-tensor field theory is consistent with conservation of charge. The next lemma provides us with the means to do just that.

**Lemma 2:** Let $A^{ab}$ and $C^a$ be the field tensor densities of a vector-tensor field theory. This theory is consistent with conservation of charge if and only if $A^{ab}$ and $C^a$ are independent of explicit dependence on $\psi_a$.

**Proof:** ⇐ Suppose that $A^{ab}$ and $C^a$ are independent of $\psi_a$; *i.e.,*

$$A^{ab;c} = 0 \text{ and } C^{a;c} = 0 \, ,$$

where

$$A^{ab;c} := \frac{\partial A^{ab}}{\partial \psi_c} \text{ and } C^{a;c} := \frac{\partial C^a}{\partial \psi_c} \, .$$

Since $A^{ab}$ and $C^a$ are the field tensor densities of a vector-tensor field theory they must

satisfy Eq.2.1, and so

$$A_a{}^b{}_{|b} = -\tfrac{1}{2} F_{ab} C^b - \tfrac{1}{2} \psi_a C^b{}_{|b}. \qquad \text{Eq.2.6}$$

Upon differentiating Eq.2.6 with respect to $\psi_a$ we get

$$0 = C^b{}_{|b},$$

which implies that charge is conserved.

⇒ Unfortunately proving the lemma in this direction will not be as easy. To begin that task let L be a Lagrangian which is such that

$$E^{ab}(L) = A^{ab} \text{ and } E^a(L) = C^a,$$

where, recall that, $C^a{}_{,a} = 0$, by assumption. We wish to prove that $A^{ab;c} = 0$, and $C^{a;c} = 0$. Since

$$C^i = - \frac{\partial L}{\partial \psi_i} + \text{a divergence}$$

we can deduce that

$$E^{ab}(C^i) = - E^{ab}\!\left[\frac{\partial L}{\partial \psi_i}\right] \text{ and } E^a(C^i) = - E^a\!\left[\frac{\partial L}{\partial \psi_i}\right]. \qquad \text{Eq.2.7}$$

Now it is easily seen that the $E^{ab}$ operator commutes with the $\frac{\partial}{\partial \psi_i}$ operator, and that the $E^a$ operator commutes with the $\frac{\partial}{\partial \psi_i}$ operator. As a result Eq.2.7 tells us that

$$E^{ab}(C^i) = - A^{ab;i} \text{ and } E^a(C^i) = - C^{a;i}. \qquad \text{Eq.2.8}$$

Hence the problem of proving that $A^{ab}$ and $C^a$ are independent of $\psi_i$ is equivalent to proving that

$$E^{ab}(C^i) = 0 \quad \text{and} \quad E^a(C^i) = 0 \qquad \text{Eq.2.9}$$

when $C^a{}_{|a} = 0$.

In passing I would like to point out equations like Eq.2.8 can be obtained from a family of differential operators associated to the Euler-Lagrange operator. These operators are discussed in Horndeski [5].

It is not too difficult to prove the validity of Eq.2.9 when $C^i$ is of arbitrary differential order. However, a few lemmas from now I shall show that when L satisfies the assumptions of the Theorem then $A^{ab}$ and $C^i$ must be at most of fourth-order in $g_{ab}$ and third-order in $\psi_a$. So let us assume that is the case. Once you see how the proof goes under those assumptions, it will be clear to you how to prove Eq. 2.9 in general. In passing I would like to mention that the future proof concerning the differential order of $A^{ab}$ and $C^a$ will not require Lemma 2 to be valid.

Under the aforementioned assumptions $E^{ab}(C^i)$ is given by

$$E^{ab}(C^i) = -C^{i;ab} + d_c C^{i;ab,c} - d_c d_d C^{i;ab,cd} + d_c d_d d_e C^{i;ab,cde} - d_c d_d d_e d_f C^{i;ab,cdef}, \quad \text{Eq.2.10}$$

where, for notational convenience, I have defined

$$C^{i;ab} := \frac{\partial C^i}{\partial g_{ab}}, \quad C^{i;ab,c} := \frac{\partial C^i}{\partial g_{ab,c}}, \quad C^{i;ab,cd} := \frac{\partial C^i}{\partial g_{ab,cd}}, \quad C^{i;ab,cde} := \frac{\partial C^i}{\partial g_{ab,cde}}, \quad C^{i;ab,cdef} := \frac{\partial C^i}{\partial g_{ab,cdef}}$$

and

$$d_c := \frac{d}{dx^c}.$$

Upon writing out the equation $C^h{}_{,h} = 0$, we get

$$0 = C^{h;rs}g_{rs,h} + C^{h;rs,t}g_{rs,th} + C^{h;rs,tu}g_{rs,tuh} + C^{h;rs,tuv}g_{rs,tuvh} + C^{h;rs,tuvw}g_{rs,tuvwh} +$$

$$+ C^{h;r}\psi_{r,h} + C^{h;r,s}\psi_{r,sh} + C^{h;r,st}\psi_{r,sth} + C^{h;r,stu}\psi_{r,stuh} ,  \qquad \text{Eq.2.11}$$

where

$$C^{h;r} := \frac{\partial C^h}{\partial \psi_r}, \quad C^{h;r,s} := \frac{\partial C^h}{\partial \psi_{r,s}}, \quad C^{h;r,st} := \frac{\partial C^h}{\partial \psi_{r,st}} \quad \text{and} \quad C^{h;r,stu} := \frac{\partial C^h}{\partial \psi_{r,stu}} .$$

Eq.2.11 is a vast reservoir of information, which I shall now tap to prove that $E^{ab}(C^i) = 0$.

To begin, since $C^h$ is fourth-order in $g_{ab}$, differentiation of Eq.2.11 with respect to $g_{ab,cdefi}$ yields

$$C^{(i;|ab|,cdef)} = 0. \qquad \text{Eq.2.12}$$

This equation tells us that the fifth term on the right-hand side of Eq.2.11 vanishes. If we act on Eq.2.12 with $d_c d_d d_e d_f$, we find that

$$d_c d_d d_e d_f \, C^{i;ab,cdef} = -4 \, d_c d_d d_e d_f \, C^{c;ab,defi} . \qquad \text{Eq.2.13}$$

Using Eq.2.13 in Eq.2.10 we find that

$$E^{ab}(C^i) = -C^{i;ab} + d_c C^{i;ab,c} - d_c d_d C^{i;ab,cd} + d_c d_d d_e \, C^{i;ab,cde} + 4 \, d_c d_d d_e d_f \, C^{c;ab,defi} . \qquad \text{Eq.2.14}$$

I shall admit that Eq.2.14 does not look like it will be that helpful, but it will be.

Now let us differentiate Eq.2.11 with respect to $g_{ab,cdei}$. The end result can be rewritten as

$$d_h C^{h;ab,cdei} + C^{(c;|ab|,dei)} = 0 . \qquad \text{Eq.2.15}$$

Differentiating Eq.2.15 with respect to $d_c d_d d_e$, and multiplying by 4 gives us

$$4\, d_c d_d d_e d_f\, C^{f;ab,cdei} + 3\, d_c d_d d_e\, C^{c;ab,dei} + d_c d_d d_e\, C^{i;ab,cde} = 0 \quad . \qquad \text{Eq.2.16}$$

Eq.2.16 permits us to rewrite Eq.2.14 as

$$E^{ab}(C^i) = -\, C^{i;ab} + d_c\, C^{i;ab,c} - d_c d_d\, C^{i;ab,cd} - 3\, d_c d_d d_e\, C^{c;ab,dei} \quad . \qquad \text{Eq.2.17}$$

So we have been able to remove the $d_c d_d d_e d_f\, C^{i;ab,cdef}$ term in $E^{ab}(C^i)$.

To continue we differentiate Eq.2.11 with respect to $g_{ab,cdi}$ to obtain

$$d_h C^{h;ab,cdi} + C^{(c;|ab|,di)} = 0 \, . \qquad \text{Eq.2.18}$$

Following our previous argument, we differentiate Eq.2.18 with respect to $d_c d_d$ and multiply by 3 to get

$$3\, d_c\, d_d\, d_e\, C^{e;ab,cdi} + 2\, d_c d_d\, C^{c;ab,di} + d_c d_d\, C^{i;ab,cd} = 0 \, . \qquad \text{Eq.2.19}$$

Eqs.2.17 and 2.19 permit us to deduce that

$$E^{ab}(C^i) = -\, C^{i;ab} + d_c C^{i;ab,c} + 2\, d_c d_d C^{c;ab,di} \quad . \qquad \text{Eq.2.20}$$

We are almost done.

Let us return to Eq.2.11 and differentiate with respect to $g_{ab,ci}$. Doing so gives us

$$d_h C^{h;ab,ci} + C^{(c;|ab|,i)} = 0 \, .$$

If we act on this equation with $d_c$ and multiply the result by 2 we find that

$$2\, d_c d_d\, C^{c;ab,di} + d_c C^{c;ab,i} + d_c C^{i;ab,c} = 0 \, .$$

Hence Eq.2.20 reduces to

$$E^{ab}(C^i) = -\, C^{ab;i} - d_c C^{c;ab,i} \, . \qquad \text{Eq.2.21}$$

The right-hand side of Eq.2.21 can be eliminated by differentiating Eq.2.11 with respect to $g_{ab,i}$. Doing so gives us

$$d_h C^{h;ab,i} + C^{i;ab} = 0 \ .$$

Therefore Eq.2.21 implies that $E^{ab}(C^i) = 0$, as required.

The proof that $E^a(C^i) = 0$, is similar, and just involves differentiating Eq.2.11 with respect to $\psi_{a,cdei}$; $\psi_{a,cdi}$; $\psi_{a,ci}$ and $\psi_{a,i}$. This task is left to the reader.

As I claimed earlier, it should be clear how to extend the proof presented above to vector-tensor field theories of arbitrary order, which are consistent with charge conservation.■

In [1] I give another proof of Lemma 2 that is only valid for second-order vector-tensor field theories. That proof is more arduous, and cannot be easily extended to higher-order vector-tensor field theories. Lemma 2 is also established in complete generality in Horndeski [6]. Nevertheless I presented the above proof to save you the trouble of locating that paper. In that work I investigate gauge invariant vector-tensor field theories, where by that I mean that the field tensor densities are invariant under the gauge transformation $\psi_a \to \psi_a + \varphi_{,a}$, where $\varphi$ is an arbitrary scalar field. In [6] it is demonstrated that gauge invariance is equivalent to charge conservation in vector-tensor field theories.

When dealing with vector-tensor field theories that are of $k^{th}$ order in the metric tensor, and $m^{th}$ order in the vector field, the derivatives of the field tensor densities, $A^{ab}$ and $C^a$, with respect to $\partial^k g_{ab}$ and $\partial^m \psi_a$ are tensorial concomitants. Here I am letting $\partial^k g_{ab}$ and $\partial^m \psi_a$ denote abbreviations for the local components of the $k^{th}$ and $m^{th}$ derivatives of $g_{ab}$ and $\psi_a$. However, in general, derivatives of $A^{ab}$ and $C^a$ with respect to $\partial^p g_{ab}$, $p<k$, and $\partial^q \psi_a$, $q<m$, are not tensorial concomitants. But under the assumptions of Lemma 2, $A^{ab;c} = 0$ and $C^{a;c} = 0$, are tensorial equations. The purpose of our next lemma is to investigate the implications of this observation.

**Lemma 3:** In an n-dimensional space if $A^{ij}$ and $C^i$ are the field tensor densities of a $k^{th}$ order vector-tensor field theory, which is consistent with charge conservation, then for every collection of s indices, a,...,b; where s = 2,..., k+1,

$$\frac{\partial A^{ij}}{\partial \psi_{(a,\ldots b)}} = 0, \text{ and } \frac{\partial C^i}{\partial \psi_{(a,\ldots b)}} = 0.$$

**Proof:** Let P be an arbitrary point in our n-dimensional space, and let x and x' denote charts at P. Since $A^{ij}$ is a tensor density we know that at P

$$A^{ij}(g'_{ab}; g'_{ab,c}; \ldots; \psi'_a; \psi'_{a,b}; \ldots) =$$

$$= |\det J^r_s| J^i_p J^j_q A^{pq}(g_{ab}; g_{ab,c}; \ldots; \psi_a; \psi_{a,b}; \ldots) \qquad \text{Eq.2.22}$$

where $J^a_b := \frac{\partial x^a}{\partial x'^b}$, $J'^i_p := \frac{\partial x'^i}{\partial x^p}$, and the derivatives of $g'_{ab}$ and $\psi'_a$ are with respect to the chart x'. Since $\psi'_a = \psi_h J^h_a$ we can deduce that for every collection of s indices a, . . . ,b; s≥2,

$$\psi'_{a,\ldots b} = \psi_h J^h_{a\ldots b} + \text{(terms involving derivatives of } \psi_a) \qquad \text{Eq.2.23}$$

where $J^h_{a\ldots b}$ denotes the $s^{th}$ derivative of $x^h$ with respect to $x'^a, \ldots, x'^b$.

In order to not get bogged down in indices here, let's assume that k = 2, in

Eq.2.22. Then upon differentiating Eq.2.22 with respect to $\psi_r$ we obtain

$$\left.\frac{\partial A^{ij}}{\partial \psi_a}\right|_{(g',\psi')} \frac{\partial \psi'_a}{\partial \psi_r} + \left.\frac{\partial A^{ij}}{\partial \psi_{a,b}}\right|_{(g',\psi')} \frac{\partial \psi'_{a,b}}{\partial \psi_r} + \left.\frac{\partial A^{ij}}{\partial \psi_{a,bc}}\right|_{(g',\psi')} \frac{\partial \psi'_{a,bc}}{\partial \psi_r} = 0 \qquad \text{Eq.2.24}$$

where $\left.\right|_{(g',\psi')}$ means to evaluate the concomitant to the left of the vertical bar for $g'_{ab}$ and $\psi'_a$. Note that when we differentiated Eq.2.22 with respect to $\psi_r$, the right-hand side vanished since $A^{ij;r} = 0$, due to Lemma 2. Similarly the first term on the left-hand side of Eq.2.24 vanishes. So if we now combine Eqs.2.23 and 2.24 we get

$$A^{ij;a,b}(g',\psi')J^r{}_{ab} + A^{ij;a,bc}(g',\psi')J^r{}_{abc} = 0 . \qquad \text{Eq.2.25}$$

The J's in Eq.2.25 are essentially arbitrary apart from the fact that $\det(J^a{}_b) \neq 0$, and $J^a{}_{b\ldots c}$ is completely symmetric in its lower indices. So if we now differentiate Eq.2.25 with respect to $J^s{}_{tu}$ and $J^s{}_{tuv}$ and then evaluate for the identity coordinate transformation, we find that at P

$$A^{ij;(t,u)} = 0 \text{ and } A^{ij;(t,uv)} = 0 . \qquad \text{Eq.2.26}$$

Since P was an arbitrary point Eq.2.26 holds in general.

The proof that $C^{i;(t,u)} = 0$ and $C^{i;(t,uv)} = 0$ is similar.

At this point it should be clear how to prove the lemma in general for arbitrary order $k \geq 2$. ∎

One might think that Lemma 3 implies that $A^{ij}$ and $C^i$ must really be built from $F_{ab}$ and its derivatives. This is in fact true, and is proved in [6]. *E.g.,* in the case where $C^i$ is fourth-order we have the replacement theorem

$$C^i = C^i(g_{ab}; \ldots ; g_{ab,cdef}; 0; \tfrac{1}{2}F_{ab}; \tfrac{2}{3}F_{a(b,c)}; \tfrac{3}{4}F_{a(b,cd)}; \tfrac{4}{5}F_{a(b,cde)}) \ .$$

However, we shall not require this fact in what follows.

The next tool we require to prove the Theorem is a generalization of a powerful identity that Aldersley developed to treat conformally invariant concomitants of the metric tensor (*see,* page 70 of [7], or [8]). To assist in its statement I shall use, as I did above, the symbols $\partial^q g_{ab}$ and $\partial^q \psi_a$ as abbreviations for the components of the $q^{th}$ derivatives of $g_{ab}$ and $\psi_a$.

**Lemma 4 (Aldersley's Identity for Conformally Invariant Vector-Tensor Field Theories):** In an n-dimensional space let

$$A^{ab} = A^{ab}(g_{hi}; \partial g_{hi}; \ldots ; \partial^k g_{hi}; \psi_h; \partial \psi_h; \ldots ; \partial^m \psi_h)$$

and

$$C^a = C^a(g_{hi}; \partial g_{hi}; \ldots ; \partial^k g_{hi}; \psi_h; \partial \psi_h; \ldots ; \partial^m \psi_h)$$

denote the field tensor densities of a conformally invariant vector-tensor field theory. Then for every real number $\lambda > 0$

$$\lambda^n A^{ab}(g_{hi}; \partial g_{hi}; \ldots ; \partial^k g_{hi}; \psi_h; \partial \psi_h; \ldots ; \partial^m \psi_h) =$$
$$= A^{ab}(g_{hi}; \lambda \partial g_{hi}; \ldots ; \lambda^k \partial^k g_{hi}; \lambda \psi_h; \lambda^2 \partial \psi_h; \ldots ; \lambda^{m+1} \partial^m \psi_h) \qquad \text{Eq.2.27}$$

and

$$\lambda^{n-1} C^a(g_{hi}; \partial g_{hi}; \ldots ; \partial^k g_{hi}; \psi_h; \partial \psi_h; \ldots ; \partial^m \psi_h) =$$
$$= C^a(g_{hi}; \lambda \partial g_{hi}; \ldots ; \lambda^k \partial^k g_{hi}; \lambda \psi_h; \lambda^2 \partial \psi_h; \ldots; \lambda^{m+1} \partial^m \psi_h) \qquad \text{Eq.2.28}$$

where there is no sum over repeated k's or m's in the arguments of $A^{ab}$ and $C^a$.

**Proof:** Once again, to make this proof more comprehensible, I shall only prove this lemma when k=m=4. From the proof presented, it will be evident how to establish the lemma in its full generality.

Let P be a point in our n-dimensional space, and let x be a chart at P. We define a new chart x' at P by $x^i = \lambda x'^i$. Since $A^{ab}$ is a tensor density we know, from the transformation law it satisfies, that at P

$$|\det(J^r_s)| J'^a_c J'^b_d A^{cd}(g_{hi}; \ldots; g_{hi,jklm}; \psi_h; \ldots; \psi_{h,ijkl}) =$$
$$= A^{ab}(g'_{hi}; \ldots; g'_{hi,jklm}; \psi'_h; \ldots; \psi'_{h,ijkl}), \qquad \text{Eq.2.29}$$

where the Jacobian matrices are defined by $J^r_s := \frac{\partial x^r}{\partial x'^s}$ and $J'^a_c := \frac{\partial x'^a}{\partial x^c}$. The tensor transformation laws for $g_{hi}$ and $\psi_h$, tell us that

$$g'_{hi} = \lambda^2 g_{hi}\,;\ g'_{hi,j} = \lambda^3 g_{hi,j}\,;\ \ldots\,;\ g'_{hi,jklm} = \lambda^6 g_{hi,jklm}\,;\ \text{and}$$

$$\psi'_h = \lambda \psi_h\,;\ \psi'_{h,i} = \lambda^2 \psi_{h,i}\,;\ \ldots\,;\ \psi'_{h,ijkl} = \lambda^5 \psi'_{h,ijkl}\,.$$

Using the above expressions in Eq.2.29, shows us that for every $\lambda > 0$, and any chart x at P,

$$\lambda^{n-2} A^{ab}(g_{hi}; \ldots; g_{hi,jklm}; \psi_h; \ldots; \psi_{h,ijkl}) =$$
$$= A^{ab}(\lambda^2 g_{hi}; \lambda^3 g_{hi,j}; \ldots; \lambda^6 g_{hi,jklm}; \lambda \psi_h; \lambda^2 \psi_{h,i}; \ldots; \lambda^5 \psi_{h,ijkl}) \,. \qquad \text{Eq.2.30}$$

I shall now show how the assumption of conformal invariance can be employed to rewrite Eq.2.30. To that end let $\gamma_{ab}$ and $\xi_a$ be the x components of a metric tensor and vector field on a neighborhood of P. Under the conformal transformation $\gamma_{ab} \to \gamma'_{ab} := \lambda^2 \gamma_{ab}$ we find that

$$A^{ab}(\lambda^2 \gamma_{hi}; \ldots; \lambda^2 \gamma_{hi,jklm}; \xi_a; \ldots; \xi_{h,ijkl}) = \lambda^{-2} A^{ab}(\gamma_{hi}; \ldots; \gamma_{hi,jklm}; \xi_h; \ldots; \xi_{h,ijkl}) \,. \text{ Eq.2.31}$$

We set

$$\gamma_{hi} = g_{hi}(P) + \lambda g_{hi,j}(P)(x^j - x^j(P)) + \tfrac{1}{2}\lambda^2 g_{hi,jk}(P)(x^j - x^j(P))(x^k - x^k(P)) +$$

$$+ \tfrac{1}{3!}\lambda^3 g_{hi,jkl}(P)(x^j - x^j(P))(x^k - x^k(P))(x^l - x^l(P)) +$$

$$+ \tfrac{1}{4!}\lambda^4 g_{hi,jklm}(P)(x^j - x^j(P))(x^k - x^k(P))(x^l - x^l(P))(x^m - x^m(P))$$

and

$$\xi_h = \lambda \psi_h(P) + \lambda^2 \psi_{h,i}(P)(x^i - x^i(P)) + \tfrac{1}{2}\lambda^3 \psi_{h,ij}(P)(x^i - x^i(P))(x^j - x^j(P)) +$$

$$+ \tfrac{1}{3!}\lambda^4 \psi_{h,ijk}(P)(x^i - x^i(P))(x^j - x^j(P))(x^k - x^k(P)) +$$

$$+ \tfrac{1}{4!}\lambda^5 \psi_{h,ijkl}(P)(x^i - x^i(P))(x^j - x^j(P))(x^k - x^k(P))(x^l - x^l(P)) .$$

Since $\gamma_{hi}(P) = g_{hi}(P)$, $\gamma_{hi}$ is a well defined metric tensor on a neighborhood of P. Using the above expressions for $\gamma_{hi}$ and $\xi_h$ in Eq.2.31 we find that at P

$$A^{ab}(\lambda^2 g_{hi}; \lambda^3 g_{hi,j}; \ldots ; \lambda^6 g_{hi,jklm}; \lambda\psi_h; \lambda^2\psi_{h,i}; \ldots ; \lambda^5 \psi_{h,ijkl}) =$$

$$= \lambda^{-2} A^{ab}(g_{hi}; \lambda g_{hi,j}; \ldots ; \lambda^4 g_{hi,jklm}; \lambda\psi_h; \lambda^2\psi_{h,i}; \ldots ; \lambda^5\psi_{h,ijkl}) . \qquad \text{Eq.2.32}$$

Upon combining Eqs.2.30 and 2.32 we discover that at P

$$\lambda^n A^{ab}(g_{hi}; g_{hi,j}; \ldots ; g_{hi,jklm}; \psi_h; \psi_{h,i}; \ldots ; \psi_{h,ijkl}) =$$

$$= A^{ab}(g_{hi}; \lambda g_{hi,j}; \ldots ; \lambda^4 g_{hi,jklm}; \lambda\psi_h; \lambda^2\psi_{h,i}; \ldots ; \lambda^5\psi_{h,ijkl}) . \qquad \text{Eq.2.33}$$

Since P was an arbitrary point, Eq.2.33 is valid in general. Eq.2.33 agrees with Eq.2.27 when k=m=4.

Using an argument similar to the one just presented for $A^{ab}$ it is easy to prove that Eq.2.28 also holds for k=m=4. Due to my earlier remarks this completes the proof of the lemma.∎

As an immediate consequence of Aldersley's identity we have

**Lemma 5:** In an n-dimensional space let

$$A^{ab} = A^{ab}(g_{hi};\, \partial g_{hi};\, \ldots;\, \partial^k g_{hi};\, \psi_h;\, \partial \psi_h;\, \ldots;\, \partial^m \psi_h)$$

and

$$C^a = C^a(g_{hi};\, \partial g_{hi};\, \ldots;\, \partial^k g_{hi};\, \psi_h;\, \partial \psi_h;\, \ldots;\, \partial^m \psi_h)$$

denote the field tensor densities of a conformally invariant, flat space compatible, vector-tensor field theory. Then $k \leq n$ and $m \leq (n-1)$ in $A^{ab}$, and $k \leq (n-1)$ and $m \leq (n-2)$ in $C^a$. In particular, in a 4-dimensional space, $k \leq 4$ and $m \leq 3$ in $A^{ab}$, and $k \leq 3$ and $m \leq 2$ in $C^a$.

**Proof:** If we differentiate Eq.2.28 with respect to $\partial^m \psi_r$ we obtain

$$\lambda^{n-1} \frac{\partial C^a}{\partial(\partial^m \psi_r)}(g_{hi};\, \partial g_{hi};\, \ldots;\, \partial^k g_{hi};\, \psi_h;\, \partial \psi_h;\, \ldots;\, \partial^m \psi_h) =$$

$$= \lambda^{m+1} \frac{\partial C^a}{\partial(\partial^m \psi_r)}(g_{hi};\, \lambda \partial_{hi};\, \ldots;\, \lambda^k \partial^k g_{hi};\, \lambda \psi_h;\, \lambda^2 \partial \psi_h;\, \ldots;\, \lambda^{m+1} \partial^m \psi_h).$$

Upon multiplying this equation by $\lambda^{(1-n)}$ we get

$$\frac{\partial C^a}{\partial(\partial^m \psi_r)}(g_{hi};\, \partial g_{hi};\, \ldots;\, \partial^k g_{hi};\, \psi_h;\, \partial \psi_h;\, \ldots;\, \partial^m \psi_h) =$$

$$= \lambda^{(m-n+2)} \frac{\partial C^a}{\partial(\partial^m \psi_r)}(g_{hi};\, \lambda \partial g_{hi};\, \ldots;\, \lambda^k \partial^k g_{hi};\, \lambda \psi_h;\, \lambda^2 \partial \psi_h;\, \ldots;\, \lambda^{m+1} \partial^m \psi_h). \qquad \text{Eq.2.34}$$

Now if $m - n + 2 \geq 1$, then we can take the limit of Eq.2.34 as $\lambda \to 0^+$. In this case the right-hand side of Eq.2.34 vanishes do to flat space compatibility. Therefore, if $m \geq n-1$, $C^a$ must be independent of $\partial^m \psi_h$. Consequently, if $C^a$ is of $m^{th}$ order in $\psi_h$, $m \leq n-2$.

In a similar way we can establish the other restrictions on k and m in $A^{ab}$ and $C^a$. ∎

Our next objective is to build all $C^a$'s that satisfy the assumptions of the Theorem. To assist in that endeavor we have

**Lemma 6:** If $C^a$ satisfies the assumptions of the Theorem then

$$C^{(a;|bc|,def)} = 0, \ C^{(a;|b|,cd)} = 0, \ C^{a;b(c,def)} = 0, \ g_{bc}C^{a;bc,def} = 0, \qquad \text{Eq.2.35}$$

$$C^{a;(b,c)} = 0 \ \text{ and } \ C^{a;(b,cd)} = 0. \qquad \text{Eq.2.36}$$

**Proof:** From Lemma 5 we know that $C^a$ is at most third-order in $g_{ab}$ and second-order in $\psi_a$. Thus the $C^a{}_{,a} = 0$ equation can be written as follows:

$$0 = C^{a;bc}g_{bc,a} + C^{a;bc,d}g_{bc,da} + C^{a;bc,de}g_{bc,dea} + C^{a;bc,def}g_{bc,defa} +$$

$$+ C^{a;b}\psi_{b,a} + C^{a;b,c}\psi_{b,ca} + C^{a;b,cd}\psi_{b,cda}. \qquad \text{Eq.2.37}$$

Upon differentiating this equation with respect to $g_{rs,tuvw}$ and $\psi_{r,stu}$ we get

$$0 = C^{(w;|rs|,tuv)} \ \text{ and } \ 0 = C^{(s;|r|,tu)},$$

which establishes the first two equations in Eq.2.35.

In passing one should note that, due to Lemma 2, the fifth term in Eq.2.37 actually vanishes.

Since $C^a$ is a contravariant vector density it must satisfy various identities, which can be obtained as follows.

Let P be an arbitrary point in our space, and let x and x' be charts at P. Due to the tensor transformation law we must have

$$C^a(g'_{hi}; g'_{hi,j}; g'_{hi,jk}; g'_{hi,jkl}; \psi'_h; \psi'_{h,i}; \psi'_{h,ij}) =$$
$$= |\det J^u_v| J'^a_r C^r(g_{hi}; g_{hi,j}; g_{hi,jk}; g_{hi,jkl}; \psi_h; \psi_{h,i}; \psi_{h,ij}),\qquad \text{Eq.2.38}$$

where the Jacobian matrices, $J^u_v$ and its inverse, $J'^a_r$, have been previously defined. At the point P

$$g'_{hi,jkl} = g_{mn}J^m_{hjkl}J^n_i + g_{mn}J^m_h J^n_{ijkl} + \text{(terms independent of } J^r_{stuv}).$$

Using this transformation equation, we discover that if we differentiate Eq.2.38 with respect to $J^r_{stuv}$, and evaluate the result for the identity transformation, we obtain

$$C^{a;hi,jkl}[g_{ri}\delta^s_{(h}\delta^t_j\delta^u_k\delta^v_{l)} + g_{hr}\delta^s_{(i}\delta^t_j\delta^u_k\delta^v_{l)}] = 0,$$

which implies that

$$C^{a;r(s,tuv)} = 0.$$

Thus we have establish the third condition in Eq.2.35.

To obtain the fourth condition let us consider the conformal transformation $g_{ab} \to g'_{ab} := e^{2\sigma}g_{ab}$. Under this transformation, since $C^a$ is conformally invariant, we find that

$$C^a((e^{2\sigma}g_{hi}); \ldots ; (e^{2\sigma}g_{hi})_{,jkl}; \psi_h; \psi_{h,i}; \psi_{h,ij}) = C^a(g_{hi}; \ldots ;g_{hi,jkl}; \psi_h; \psi_{h,i}; \psi_{h,ij}). \quad \text{Eq.2.39}$$

If we differentiate Eq.2.39 with respect to $\sigma_{,rst}$, and then evaluate the result for the identity conformal transformation we obtain

$$C^{a;hi,rst} g_{hi} = 0.$$

This completes our proof of the validity of Eq.2.35. Eq.2.36 follows from

Lemma 3.■

At last we are ready to determine the basic functional form of $C^a$. This will be done in the next lemma.

**Lemma 7:** If $C^a$ satisfies the assumptions of the Theorem then

$$C^a = \Theta^{abcdef}g_{bc,def} + \Theta^{abcdefhi}g_{bc,de}g_{fh,i} + \Theta^{abcdefhijk}g_{bc,d}g_{ef,h}g_{ij,k} +$$

$$+ \Psi^{abcd}\psi_{b,cd} + \Psi^{abcdef}\Psi_{b,c}g_{de,f} \qquad \text{Eq.2.40}$$

where the $\Theta$'s and $\Psi$'s are concomitants of only $g_{ab}$. $\Theta^{abcdef}$ and $\Psi^{abcd}$ are tensorial concomitants which have the following symmetries:

$$\Theta^{abcdef} = \Theta^{a(bc)def} = \Theta^{abc(def)}, \ \Theta^{ab(cdef)} = 0, \ \Theta^{(a|bc|def)} = 0, \ g_{bc}\Theta^{abcdef} = 0; \quad \text{Eq.2.41}$$

and

$$\Psi^{abcd} = \Psi^{ab(cd)}, \ \Psi^{(a|b|cd)} = 0, \ \Psi^{a(bcd)} = 0 \ . \qquad \text{Eq.2.42}$$

**Proof:** Due to Lemma 2, Aldersley's identity (Lemma 4), and Lemma 5, we know that for every $\lambda > 0$

$$\lambda^3 C^a(g_{hi}; \ldots ; g_{hi,jkl}; \psi_{h,i}; \psi_{h,ij}) = C^a(g_{hi}; \lambda g_{hi,j}; \lambda^2 g_{hi,jk}; \lambda^3 g_{hi,jkl}; \lambda^2 \psi_{h,i}; \lambda^3 \psi_{h,ij}) \ . \qquad \text{Eq.2.43}$$

Upon differentiating this equation with respect to $g_{rs,tuv}$ we find that

$$C^{a;rs,tuv}(g_{hi}; \ldots ; g_{hi,jkl}; \psi_{h,i}; \psi_{h,ij}) =$$

$$= C^{a;rs,tuv}(g_{hi}; \lambda g_{hi,j}; \lambda^2 g_{hi,jk}; \lambda^3 g_{hi,jkl}; \lambda^2 \psi_{h,i}; \lambda^3 \psi_{h,ij}) \ . \qquad \text{Eq.2.44}$$

If we differentiate this equation with respect to $g_{hi,jkl}$, and then take the limit as $\lambda \to 0^+$, recalling that $C^a$ is well defined and differentiable when evaluated for a flat metric tensor and vanishing vector field, we see that

$$C^{a;rs,tuv;hi,jkl} = 0 \ .$$

Similarly we can use Eq.2.44 to prove that

$C^{a;rs,tuv;hi,jk} = 0$, $C^{a;rs,tuv;hi,j} = 0$, $C^{a;rs,tuv;h,ij} = 0$, and $C^{a;rs,tuv;h,i} = 0$ .

Consequently $g_{bc,def}$ must appear linearly in $C^a$, with coefficients that are solely functions of $g_{ab}$.

Analogously we can demonstrate that $\psi_{b,cd}$ appears linearly in $C^a$ with coefficients that depend only on $g_{ab}$.

Continuing in this fashion we can use Eq.2.43 to show that $C^a$ must be a linear combination of $g_{bc,def}$ ; $g_{bc,de}g_{fh,i}$ ; $g_{bc,d}g_{ef,h}g_{ij,k}$ ; $\psi_{b,cd}$ ; and $\psi_{b,c}g_{de,f}$ ; with coefficients which are simply functions of $g_{rs}$.

The symmetries satisfied by $\Theta$ and $\Psi$ in Eqs.2.41 and 2.42 follow from Lemma 6, along with the symmetries inherent in partial derivatives with respect to $g_{bc,def}$ and $\psi_{b,cd}$. ∎

In order to simplify the form of $C^a$ given in Lemma 7, we need

**Lemma 8 (Thomas's Replacement theorem for Vector-Tensor Concomitants):**
If $\tau$ is a tensorial concomitant which locally has the form

$$\tau^{\cdots}_{\cdots} = \tau^{\cdots}_{\cdots}(g_{hi}; g_{hi,j}; g_{hi,jk}; g_{hi,jkl}; \psi_h; \psi_{h,i}; \psi_{h,ij}) ,$$

then the value of $\tau^{\cdots}_{\cdots}$ is unaffected if its arguments are replaced as shown below

$$\tau^{\cdots}_{\cdots} = \tau^{\cdots}_{\cdots}(g_{hi}; 0; \tfrac{1}{3}(R_{hjki} + R_{hkji}); \tfrac{1}{6}(R_{hjki|l} + R_{hkji|l} + R_{hkli|j} + R_{hlki|j} + R_{hlji|k} + R_{hjli|k});$$

$$\psi_h; \psi_{h|i}; \psi_{h|(ij)} + \tfrac{1}{6}\psi_m(R_i{}^m{}_{hj} + R_j{}^m{}_{hi})) .$$

**Proof:** I essentially explain why Thomas's Replacement Theorem [9], is valid in [2],

by simply reformulating his arguments in Appendix B of [2]. Nevertheless I shall quickly sketch how it works because of the terms involving the derivatives of $\psi_h$.

Let P be an arbitrary point in our space, and let x be a chart at P. x gives rise to a normal coordinate system y at P which is such that $\frac{\partial}{\partial x^i} = \frac{\partial}{\partial y^i}$ at P, and hence $\frac{\partial x^i}{\partial y^j} = \delta^i_j$ at P. Let $\gamma_{ab}$ and $\xi_a$ denote the y components of the metric tensor and vector field. Due to the fact that $\tau^{\cdots}_{\cdots}$ are the components of a tensorial concomitant we know that the x and y components of $\tau$ at P must be related by

$$\tau^{\cdots}_{\cdots}(g_{hi}; \ldots; g_{hi,jkl}; \psi_h; \psi_{h,i}; \psi_{h,ij}) = \tau^{\cdots}_{\cdots}(\gamma_{hi}; \ldots; \gamma_{hi,jkl}; \xi_h; \xi_{h,i}; \xi_{h,ij}) , \qquad \text{Eq.2.45}$$

where the derivatives on the left-hand side of Eq.2.45 are taken with respect to the chart x, and those on the right-hand side are taken with respect to the chart y. $\gamma_{hi}; \ldots, \gamma_{hi,jkl}; \xi_h; \xi_{h,i}$ and $\xi_{h,ij}$ are the componenets of tensors at P, called the extensions of $g_{ab}$ and $\psi_a$. On page 99 of [9] Thomas gives formulas for these tensor components in terms of their components with respect to the arbitrary chart x at P. When these expressions are placed into Eq.2.45, we obtain the proof of our lemma.

If you do not have access to Thomas's book, you can derive all of the formulas for the extensions of $g_{ab}$ and $\psi_a$ we require using Appendix B in [2]. The formulas you will obtain will involved the Christoffel symbols of the second kind and their derivatives. The key to rewriting those expressions in terms of more familiar quantities is to evaluate them at the pole of a normal coordinate system. *E.g.,* at the pole, the second extension of $\psi_h$, denoted $\psi_{h;ij}$, is given by

$$\psi_{h;ij} = \psi_{h,ij} - \psi_m \Gamma^m{}_{hij},$$

where, at the pole,

$$\Gamma^m{}_{hij} = \tfrac{1}{3}(\Gamma^m{}_{hi,j} + \Gamma^m{}_{ij,h} + \Gamma^m{}_{jh,i}).$$

Upon combining these two expressions we see that

$$\psi_{h;ij} = \psi_{h|(ij)} + \tfrac{1}{6}\psi_m(R_i{}^m{}_{hj} + R_j{}^m{}_{hi}).$$

$\psi_{h;ij}$ is what replaces $\psi_{h,ij}$ in $\tau^{\cdots}_{\cdots}$. ∎

Due to Thomas's Replacement Theorem we see that Eq.2.40 reduces to

$$C^a = \Theta^{abcdef} R_{bdec|f} + \Psi^{abcd}(\psi_{b|(cd)} + \tfrac{1}{3}\psi_m R_c{}^m{}_{bd}), \qquad \text{Eq.2.46}$$

where I have made use of the symmetries of $\Theta$ and $\Psi$.

At first sight you might think that Eq.2.46 must be wrong since Lemma 2 tells us that $C^a$ must be devoid of explicit dependence on $\psi_a$. But we need to know something about $\Psi^{abcd}$ before we start to panic. This is where our next lemma comes to the rescue.

**Lemma 8:** If $\Theta^{abcdef}$ and $\Psi^{abcd}$ are tensorial concomitants of $g_{ab}$ which satisfy Eq.2.41 and 2.42, then

$$\Theta^{abcdef} = 0 \text{ and } \Psi^{abcd} = g^{½} \alpha(g^{ab}g^{cd} - \tfrac{1}{2}g^{ac}g^{bd} - \tfrac{1}{2}g^{ad}g^{bc}), \qquad \text{Eq.2.47}$$

where $\alpha$ is a constant.

**Proof:** In Appendix C of [2] it is shown how some of the results Weyl presents in [10] can be used to construct $\Theta^{abcdef}$ and $\Psi^{abcd}$. The main result we require is that $\Theta^{abcdef}$ can be built from a linear combination of terms built from the product of three $g^{..}$'s, while $\Psi^{abcd}$ can be constructed from a linear combinations of terms involving two g..'s, where, in both cases, the coefficients in the linear combinations are constants. (Terms of the form $g^{ab}\varepsilon^{cdef}$, where $\varepsilon^{cdef}$ is the Levi-Civita tensor density, cannot appear in $\Theta^{abcdef}$ because of its symmetry in the indices b,c and d,e,f.) If you express $\Theta^{abcdef}$ as a linear combination of terms involving three g..'s, you will start with an ansatz expression involving 15 terms. Upon imposing all of $\Theta^{abcdef}$'s symmetries on this initial candidate you easily find that it must vanish.

Now for $\Psi^{abcd}$, your initial expression will look like this

$$\Psi^{abcd} = g^{\frac{1}{2}}(\alpha g^{ab}g^{cd} + \beta g^{ac}g^{bd} + \gamma g^{ad}g^{bc}),$$

where $\alpha$, $\beta$ and $\gamma$ are constants. After imposing the symmetries that $\Psi^{abcd}$ enjoys upon this expression we find that $\Psi^{abcd}$ is indeed given by Eq.2.47.∎

Our next lemma provides us with our long sought expression for $C^a$.

**Lemma 9:** If $C^a$ satisfies the assumptions of the Theorem then

$$C^a = \alpha\, g^{\frac{1}{2}}\, F^{ab}{}_{|b}, \qquad \text{Eq.2.48}$$

where $\alpha$ is a constant. A Lagrangian which yields $C^a$ as its Euler-Lagrange tensor upon varying the vector field is $\beta L_M$, where $L_M$ is defined by Eq.1.6 and $\beta := -\frac{1}{4}\alpha$.

**Proof:** Under the assumptions of the Theorem it was shown that $C^a$ must have the form given in Eq.2.46. Thus due to Lemma 8 we can conclude that

$$C^a = g^{\frac{1}{2}}\alpha(g^{ab}g^{cd} - \tfrac{1}{2}g^{ac}g^{bd} - \tfrac{1}{2}g^{ad}g^{bc})(\psi_{b|cd} + \tfrac{1}{3}\psi_m R_c{}^m{}_{bd}). \qquad \text{Eq.2.49}$$

Upon multiplying out the right-hand side of Eq.2.49 we find that

$$C^a = \tfrac{1}{2} g^{\frac{1}{2}} \alpha (F^{ac}{}_{|c} + \psi^{a|c}{}_{|c} - \psi^{c}{}_{|c}{}^{|a} - \psi_m R^{am}) .  \qquad \text{Eq.2.50}$$

However,

$$\psi^{c}{}_{|c}{}^{|a} = \psi^{c|a}{}_{|c} - \psi^m R_m{}^a ,$$

and therefore Eq.2.50 reduces to

$$C^a = \alpha\, g^{\frac{1}{2}}\, F^{ab}{}_{|b} ,$$

as required.

The fact that $C^a$ can be derived from the Lagrangian $\beta L_M$, when $\beta = -\tfrac{1}{4}\alpha$, follows from the remarks in Section 1. ∎

I can not help but remark upon the amazing logical consistency of mathematics which we just witnessed. Thomas's Replacement Theorem knew nothing about the implications of charge conservation when it told us to replace $\psi_{h,ij}$ by $\psi_{h|(ij)} + \tfrac{1}{6}\psi_m(R_i{}^m{}_{hj} + R_j{}^m{}_{hi})$. And yet the demands of charge conservation dictated the coefficients of $\psi_{h,ij}$ in $C^a$ have just the right symmetries to eliminate the $\psi_m$ terms.

We are now poised to complete the proof of the Theorem. To that end let L be a Lagrangian which yields a vector-tensor field theory satisfying the assumptions of the Theorem. We define $L := L - \beta L_M$, where due to Lemma 9 we know that we can choose $\beta$ so that $E^a(L) = 0$. The purpose of our next lemma is to determine a pure metric Lagrangian equivalent to $L$ from a variational point of view.

**Lemma 10:** Suppose that in a 4-dimensional space the $k^{th}$ order Lagrangian $L$ generates a conformally invariant, flat space compatible, vector-tensor field theory for which $E^a(L) = 0$. Then $E^{ab}(L) = E^{ab}(bL_B)$ for a suitable choice of the constant b, where $L_B$ is the Bach Lagrangian defined by Eq.1.10.

**Proof:** Let us consider the one-parameter variation of $\psi_a$ defined by $\psi(t)_a := t\,\psi_a$, $0 \leq t \leq 1$. Correspondingly we define the one-parameter family of Lagrangians $L(t)$ by

$$L(t) := L(g_{ab}; \partial g_{ab}; \ldots; \partial^k g_{ab}; \psi(t)_a; \partial \psi(t)_a; \ldots; \partial^k \psi(t)_a) \,. \qquad \text{Eq.2.51}$$

Note that since $L$ defines a flat-space compatible Lagrangian, $L(0)$ is well-defined. If we now use the usual variational arguments, we find that since $E^a(L(t)) = 0$,

$$\frac{dL(t)}{dt} = V(t)^i{}_{,i} \qquad \text{Eq.2.52}$$

where $V(t)^i$ is a one-parameter family of contravariant vector fields. Upon integrating Eq.2.52 with respect to t from 0 to 1 we get

$$L(1) - L(0) = \text{a divergence.} \qquad \text{Eq.2.53}$$

Eq.2.51 tells us that $L(1) = L$, while $L(0)$ is a pure metric Lagrangian. So Eq.2.53 implies that the field theory generated by $L$ can also be generated by a pure metric Lagrangian which is flat space compatible, and gives rise to conformally invariant field tensor densities. In the Corollary presented in [2], it is shown that in a 4-dimensional space, any conformally invariant pure metric theory, which is flat space compatible, can have its field tensor densities generated by the Lagrangian $bL_B$, for a suitable choice of the constant b. This observation completes the proof of the Lemma.∎

Due to Lemmas 9 and 10, we know that if L is a Lagrangian that satisfies the assumptions of the Theorem, then there exists a constant β, for which, $L - \beta L_M$, generates a theory that could be obtained from the Lagrangian $bL_B$ for a suitable choice of the constant b. Thus the theory generated by L can also be generated by $bL_B + \beta L_M$. This is precisely what we have been trying to demonstrate.

It should be noted that Lemma 10 marks the first, and only time, that I used the assumption that L was defined and differentiable for a vanishing vector field, in the proof of the Theorem. I believe that the Theorem can be proved if we replace the current assumption of flat space compatiblity by the demand that the field tensor densities be defined and differentiable for either a flat metric tensor (and) or vanishing vector field. However, proving the Theorem under these weaker assumptions is much more difficult. What we would have to do is actually construct $A^{ab} := E^{ab}(L)$, when $E^a(L) = 0$. To that end you can use Aldersley's identity to get the basic form of $A^{ab}$, as we did for $C^a$, in Lemma 7. Then your task would be to prove that all terms involving the derivatives of $\psi_a$ in $A^{ab}$ vanish. You can use the fact that $E^c(A^{ab}) = 0$, and $A^{ab}{}_{|b} = 0$, to help in that endeavor. Once this is accomplished you can appeal to the Corollary in [2] to finish the proof of this new and improved version of the Theorem.

**Section 3: Concluding Remarks**

In [2] I constructed all of the conformally invariant, flat space compatible,

scalar-tensor field theories, and it turned out that there were four distinct classes of such theories, all of which had arbitrary functions of the scalar field appearing in them. So when I originally began examining the vector-tensor problem of finding all conformally invariant, flat space compatible, vector-tensor field theories consistent with charge conservation, I expected to find more than one class of true vector-tensor field theories. Well, my expectations were pleasantly dashed, since we now know that the only true vector-tensor field theory that satisfies the aforementioned assumptions can be generated by a constant multiple of the Maxwell Lagrangian. And moreover, the only possible vector field equation is Maxwell's.

These observations go a long way toward answering Einstein's question of whether God (if it exists) had any choice, when it came to selecting field equations to govern our Universe. For due to Lovelock's work [11] we know that if you want second-order metric field equations to describe regions devoid of matter, and you desire these equations to come from a variational principle, then those equations can be derived from $L = g^{½}R - g^{½}\Lambda$, where $\Lambda$ is the cosmological constant. Now if you want to add an electromagnetic source to the gravitational field equations, and you want that source to be obtained from a flat space compatible, conformally invariant, true vector-tensor field theory, which is consistent with charge conservation, then there is only one other Lagrangian you can add to Lovelock's: *viz.,* $\beta L_M$, a constant multiple of the Maxwell Lagrangian. We would then choose $\beta = 1$ to get the

"numbers right."

A generalization of the Einstein-Maxwell field equations is provided by the Einstein-Yang-Mills field equations, which present us with an example of a gauge-tensor field theory. (*See,* Yang, Mills [12] for a discussion of Yang-Mills theory.) If we let $\psi^\alpha_i$ denote the gauge potentials (where small Greek indices run from 1 to n, and n is the dimension of the gauge group G, which is a Lie group), then the components of its associated curvature tensor are given by $F^\alpha_{ij} := \psi^\alpha_{i,j} - \psi^\alpha_{j,i} - C^\alpha_{\beta\gamma} \psi^\beta_i \psi^\gamma_j$, where $C^\alpha_{\beta\gamma}$ denotes the structure constants of the Lie algebra, LG, of G. (I realize that my Greek and Latin indices are just the opposite of those conventionally used, but they are consistent with my previous index usage.) A Lagrangian that yields the Einstein-Yang-Mills field equations is given by

$$L_{EYM} := g^{\frac{1}{2}} R + L_{YM} \qquad \text{Eq.3.1}$$

where

$$L_{YM} := g^{\frac{1}{2}} B_{\alpha\beta} F^\alpha_{ij} F^{\beta ij}, \qquad \text{Eq.3.2}$$

and $B_{\alpha\beta}$ denotes the components of a symmetric AdG invariant bilinear form on LG. (By $B_{\alpha\beta}$ being AdG invariant I mean that for every $h \in G$, $B_{\alpha\beta} = B_{\mu\nu} Ad^\mu_\alpha(h) Ad^\nu_\beta(h)$.) The Yang-Mills Lagrangian, given in Eq.3.2, is conformally invariant, and generates a gauge-tensor field theory which is consistent with (gauge) charge conservation in that $E^a_\alpha(L_{YM})_{||a} = 0$, where $E^a_\alpha(L_{YM})_{||a} := E^a_\alpha(L_{YM})_{,a} - E^a_\beta(L_{YM}) C^\beta_{\gamma\alpha} \psi^\gamma_a$. We recover the Einstein-Maxwell theory from this gauge theory by choosing the Lie group G to be ℝ, and then $B_{\alpha\beta}$ has only one entry, which we take to be $-1$.

Now I believe that it should be possible to modify the theory developed in section 2 using the material presented in Horndeski [13] to establish the following

**Conjecture 1:** In a space of four-dimensions let L be a Lagrangian which generates a conformally invariant, flat space compatible, gauge-tensor field theory which is consistent with charge conservation. Then the Euler-Lagrange tensors associated with L can also be obtained from the Lagrangian $bL_B + L_{YM}$, where b is a constant, with $L_B$ and $L_{YM}$ defined by Eqs.1.10 and 3.2.∎

If this conjecture turns out to be true, it would provide us with another reason to demand that the field equations associated with source fields should be required to be conformally invariant, since that demand leads to unique field equations in several instances.

However, I have to mention that I have been a little surreptitious here. For if we were to assume that our manifold is orientable, then we could have included the conformally invariant Lagrangian

$$L_{YM*} := B_{\alpha\beta}\, \varepsilon^{hijk} F^\alpha_{hi}\, F^\beta_{jk}$$

in the statement of the conjecture. Note that $L_{YM*}$ becomes a divergence in the abelian case, and that is why no such term appears in the Theorem.

Let us now return to vector-tensor field theory. Just because Maxwell's equations work very well in flat space, does not imply that they should be the equations we use in curved space. That is why the result I present in [1] is significant. Since it essentially shows that there is only one second-order vector-tensor alternative to Maxwell's theory in a curved space, that is consistent with charge conservation, and reduces to Maxwell's equations in flat space.

In view of what was accomplished in Section 2, is there anything we can say about vector-tensor field theories in a four-dimensional space, that are conformally invariant, and flat space compatible, but not necessarily consistent with charge conservation? It turns out that we can actually say quite a lot. For let L be the Lagrangian of such a theory. If both sets of field tensor densities generated by L do not depend explicitly on $\psi_a$, we can use Lemma 2 to deduce that L generates a field theory compatible with conservation of charge. Thus we know what L is equivalent to in that case, due to our Theorem. So let us assume that at least one of the field tensor densities generated by L has explicit $\psi_a$ dependence. Then we define $L_\varphi$ to be the scalar-tensor Lagrangian we get from L by replacing $\psi_a$ and its derivatives, by $\varphi_{,a}$ and its derivatives, where $\varphi$ is a scalar field. Since L yielded a conformally invariant vector-tensor field theory, $L_\varphi$ will yield a conformally invariant scalar-tensor field theory. In [2] I constructed all flat space compatible, conformally invariant, scalar-tensor field theories in an orientable four-dimensional space. From that work we know that the field tensor densities derivable from $L_\varphi$ can also be obtained from the Lagrangian

$$L_C := L_{2C} + L_{3C} + L_{4C} + L_{UC},$$

where, in the present case,

$$L_{2C} = g^{½} k\, (\varphi_{,a}\varphi_{,b}\, g^{ab})^2,$$

$$L_{3C} = p\, \varepsilon^{abcd} C^{pq}{}_{ab} C_{pqcd},$$

$$L_{4C} = b\, L_B,$$

and

$$L_{UC} = g^{½} u(-12 R^{ab}\varphi_{,a}\varphi_{,b} + 2R\varphi_{,a}\varphi_{,b} g^{ab} - 3(\Box\varphi)^2 - 6\varphi^{|ab}\varphi_{|ab} - 12\varphi^{,a}(\Box\varphi)_{,a}),$$

where k, p, b and u are constants, and $L_B$ is defined by Eq.1.10.

Evidently, when the substitution $\psi_a \to \varphi_{,a}$ is made, the Lagrangian $L_{2C}$ can be obtained from the conformally invariant, flat space compatible, vector-tensor Lagrangian $k\, L_{2CV}$, where

$$L_{2CV} := g^{½}(\psi_a\, \psi_b\, g^{ab})^2. \qquad \text{Eq.3.3}$$

Thus we have $k\, L_{2CV\varphi} = L_{2C}$.

The Lagrangian $L_{3C}$ is trivial in the present case, since its Euler-Lagrange tensor densities vanish. $L_{4C}$ corresponds to a constant multiple of the Bach Lagrangian given in Eq.1.10, which we know to generate a theory consistent with charge conservation.

Associated with the Lagrangian $L_{UC}$, is the vector-tensor Lagrangian $L_{UCV}$, defined by

$$L_{UCV} := g^{½} u(-12 R^{ab}\psi_a\psi_b + 2R\psi_a\psi_b g^{ab} - 3(\psi^a{}_{|a})^2 - 6\psi^{a|b}\psi_{a|b} - 12\psi^a(\psi^b{}_{|b})_{,a}).$$

Although $L_{UCV}$ has the property that, $L_{UCV\varphi} = L_{UC}$, it is not conformally invariant, nor is it conformally invariant up to a divergence. The problem term is $\psi^{a|b}\psi_{a|b}$. Replacing it with $\psi^{a|b}\psi_{b|a}$ or $\frac{1}{2}(\psi^{a|b} + \psi^{b|a})\psi_{a|b}$ does not help to yield a conformally invariant vector-tensor field theory.

The upshot of the above analysis is that if the flat space compatible Lagrangian L, yields a conformally invariant vector-tensor field theory, then the field theory it generates when we make the substitution $\psi_a = \varphi_{,a}$ can also be generated by the vector-tensor Lagrangian $kL_{2CV} + \beta L_M + bL_B$, for a suitable choice of the constants k, β and b, when we make the same substitution $\psi_a = \varphi_{,a}$. However, this does not imply that in general L yields the same vector-tensor field equations as does $L_{2CV} + \beta L_M + bL_B$. For there might be a conformally invariant vector-tensor Lagrangian, Λ, which is such that $\Lambda_\varphi$ either vanishes, or is a divergence, and the field tensor densities generated by Λ depend explicitly on $\psi_a$. (*E.g.*, when working in a six-dimensional space we can take Λ to be $\Lambda := g^{\frac{1}{2}}g^{ab}g^{cd}g^{ef}\psi_a \psi_c F_{be} F_{df}$.) Nevertheless, I am willing to make the following

**Conjecture 2:** In a space of four-dimensions the field tensor densities of any flat space compatible, conformally invariant, vector tensor field theory can be derived from the Lagrangian $kL_{2CV} + \beta L_M + bL_B$ for a suitable choice of the constants k, b and β. The Lagrangians $L_{2CV}$, $L_M$ and $L_B$ are defined by Eqs.3.3, 1.6 and 1.10.■

This paper is the second of my trilogy of papers dealing with conformally invariant field theories. In the next, and final paper, I shall discuss conformally invariant scalar-vector-tensor field theories, that are flat space compatible and consistent with conservation of charge.

**Acknowledgements**

I would like to thank Dr.M.Zumalacárregui**,** and Dr.M.Crisostomi for discussions relating to conformally invariant field theories.